\documentclass[reprint,aip,amsmath,amssymb,jcp,superscriptaddress]{revtex4-2}

\PassOptionsToPackage{letterpaper,top=2cm,bottom=2cm,left=3cm,right=3cm,marginparwidth=1.75cm}{geometry}

\usepackage[english]{babel}

\usepackage{geometry}

\usepackage{amsmath,bm}
\usepackage{graphicx}
\usepackage[colorlinks=true, allcolors=blue]{hyperref}
\usepackage{amssymb}
\usepackage{mathrsfs}
\usepackage{xcolor}
\usepackage{xfrac}
\usepackage{physics}
\usepackage{comment}
\usepackage{float}
\usepackage{subcaption}
\usepackage{bm}
\usepackage[cal=dutchcal]{mathalfa}

\usepackage{stfloats}    
\newcommand{\hH}{{\hat{H}}}

\newcommand{\bGamma}{{\bm{\hat{\Gamma}}}}
\newcommand{\bd}{{\bm{d}}}
\newcommand{\bR}{{\bm{R}}}
\newcommand{\bP}{{\bm{P}}}
\newcommand{\bK}{{\bm{K}}}

\newcommand{\br}{{\bm{r}}}
\newcommand{\bp}{{\bm{p}}}

\newcommand{\bx}{{\bm x}}

\newcommand{\bJ}{{\bm J}}

\newcommand{\bM}{{\bm M}}

\begin{document}

\title{Conical Intersections and  Electronic Momentum As Viewed From Phase Space Electronic Structure Theory}

\begin{abstract}
    We investigate the structure of a prototypical two-state conical intersection (BeH$_2$) using a phase space electronic Hamiltonian $\hat{H}_{PS}(\bR,\bP)$ that goes beyond the Born-Oppenheimer framework. By parameterizing the electronic Schr{\"o}dinger equation by both nuclear position ($\bR$) and momentum ($\bP$), we solve for quantum electronic states in a moving frame that can break time reversal symmetry and, as a result, the branching plane of the conical intersection within a phase space framework now has dimension three (rather than dimension two as found within the standard Born-Oppenheimer framework). Moreover, we note that, if one fixes a geometry in real space that lies in the conical intersection seam and scans over the corresponding momentum space, one finds a double well (with minima at $\pm \bP_{min} \ne 0$), indicating that the stationary electronic states of the phase space electronic Hamiltonian carry electronic momentum -- a feature that cannot be captured by a Born-Oppenheimer electronic state. Interestingly, for $BeH_2$, this electronic momenta (as calculated with full configuration interaction) agrees with what is predicted by approximate complex restricted Hartree-Fock calculations, indicating a physical interpretation of complex instabilities in modern electronic structure calculations. Altogether, this study suggests that we have still have a lot to learn about conical intersections when it comes to electronic momentum, and highlights the urgent need for more experiments to probe what photochemical observables can and/or cannot be captured by standard electronic structure that isolates conical intersections within the Born-Oppenheimer framework. 
    
\end{abstract}

\author{Titouan Duston}
\affiliation{Department of Chemistry, Princeton University, Princeton, New Jersey 08540, USA.}
\author{Nadine Bradbury}
\affiliation{Department of Chemistry, Princeton University, Princeton, New Jersey 08540, USA.}
\author{Zhen Tao}
\affiliation{Department of Chemistry, Princeton University, Princeton, New Jersey 08540, USA.}
\author{Joseph E. Subotnik}
\email{subotnik@princeton.edu}
\affiliation{Department of Chemistry, Princeton University, Princeton, New Jersey 08540, USA.}

\maketitle

\section{Introduction: Electronic Momentum in Nonadiabatic Processes}

There are many fundamental processes in chemistry that require advanced nuclear-electronic dynamics methods that allow for motion on multiple electronic surfaces. In particular, tractable and accurate nonadiabatic techniques for dynamics on multiple surfaces are required to study
photochemistry\cite{C2H4photochem_Martinez,Yarkony_photoCI2012}, electron transfer\cite{ET_nonadiabat}, and inter-system crossings.\cite{ISCreview} 
To that end, great strides have been taken over the last few decades to develop efficient nonadiabatic algorithms and nowadays it is possible to navigate somewhat successfully on multiple surfaces. That being said, it must be noted that computational chemistry today still remains tied to the Born-Oppenheimer (BO) {\em framework} insofar as,
with only a few exceptions\cite{truhlar:review:surfacehop},
standard nonadiabatic approaches (e.g., fewest switches surface hopping\cite{Tully1990, Tully1998} and multiple spawning\cite{AIMS_martinez, AIMS_martinez_photo}) 
run along BO surfaces. Linear response Ehrenfest dynamics\cite{ehrenfest_2005} is also tied to the BO representation.
In short, while theoretical chemists have long sought to go beyond the BO {\em approximation} and have achieved many successes in that quest over the past several decades, the fact remains that the vast majority of theoretical chemistry remains driven by the BO {\em framework}.

As far as understanding the strengths and limitations of the above algorithms to solve nonadiabatic problems, most analysis heretofore has focused on understanding the {\em energy} flow between nuclei and electrons. 
For instance, the surface hopping and multiple spawning algorithms were designed to capture how electronic transitions absorb or release energy to and from the nuclear degrees of freedom. 
The redistribution of energy is achieved within Ehrenfest dynamics by walking along average potential energy surfaces. In order to quantify the flow of energy between nuclei and electrons, chemists nowadays routinely calculate
branching ratios\cite{Tully1990,zheng2009phase,cotton:2016:et_and_tully:jcp} from scattering scenarios and kinetics (e.g. Marcus theory rates\cite{marcus1985electron,landry:2011:marcus_fssh,reichman:2016:fssh,jain:2015:friction}). There has also been a focus on the final equilibrium state\cite{tully:2005:detailedbalance,tully:2008:detailedbalance,tully2012perspective} as predicted by different nonadiabatic schemes. For example, it is well known that Ehrenfest does not obey detailed balance, whereas surface hopping does (but only approximately\cite{barbatti2011nonadiabatic}). 


Now, while the focus above is on energy conversion, 
one missing ingredient in the discussion above is momentum: the redistribution of momentum between nuclei and electrons is rarely discussed in the literature (except within the community of exact-factorization theorists\cite{gross2010exactfactor,li_gross:2022:PRL:angular_momentum_transfer}). Ultimately, the source of the problem is clear: by making a separation between nuclei and electrons, and calculating electronic orbits in the frame of frozen nuclei, BO theory does not treat electronic and nuclear momentum equivalently. More precisely, while BO theory conserves momentum with the proper interpretation\cite{littlejohn:2023:jcp:angmom}, in practice, momentum conservation is often achieved by setting the electronic momentum to zero. Therefore, one cannot reliably calculate the flow of momentum between nuclei and electrons (though the total should obviously be conserved); for meaningful momentum-conserving dynamics within a BO framework, one must include non-BO terms.\cite{xuezhi:2023:total_ang_bomd,coraline:2024:jcp:ehrenfest_conserve} At this point, in order to understand the implications of this error in detail, let us analyze two separate cases in BO theory: the cases of dynamic vs. static coupling.


\subsection{The Limit of Dynamic Correlation}

Consider first the case where a molecule evolves on one BO surface that is well separated from the other surfaces. For motion on a single surface (here labeled $I$), it is obvious that there can be no electronic momentum within BO theory. After all, $\left< \Phi_I \middle| \hat{\bp}_e \middle| \Phi_I \right> = 0$ for any real electronic wavefunction $\ket{\Phi_I}$.

Of course, the result above is incomplete. A more accurate result was found by Nafie, who used perturbation theory.\cite{nafie1983} The argument is as follows: 
If we regard the term $- i\hbar \frac{\bP \cdot \hat{\bm d}}{\bM}$ as the perturbation to the standard $H_{el}$ within the BO framework, then we can construct the perturbed electronic wavefunction $\ket{\Psi_I}$ as

\begin{eqnarray}
\label{eq:psi}
    \ket{\Psi_I} = \ket{\Phi_I} - i\hbar \sum_{J\ne I}\frac{ \langle \Phi_J | \sum_A \frac{\bP^A \cdot \hat{\bd}^A}{M_A}  |\Phi_I\rangle}{E_I -E_J} \ket{\Phi_J}
\end{eqnarray}
where $\left< \Phi_I \middle| \hat{\bm d}^A \middle| \Phi_J \right> \equiv \bm d^A_{IJ}$ is the derivative coupling vector. Evaluating the expectation value of $\langle \Psi_I |\hat{\bp}_e |\Psi_I\rangle$, we find:

\begin{eqnarray}
    &&\langle \Psi_I|\hat{\bp}_e | \Psi_I \rangle =\\ &&2 \hbar \text{Im}  \sum_{J\ne I} \langle \Phi_I |\hat{\bp}_e | \Phi_J\rangle\frac{ \langle \Phi_J | \sum_A \frac{\bP^A \cdot \hat{\bd}^A}{M_A}  |\Phi_I\rangle}{E_I -E_J} \label{eq:pe_nafie}
\end{eqnarray}

\noindent If we assume a complete basis, we may use the commutator:

\begin{eqnarray}
    \left[ \hH_{e},\hat{\br}_e \right] = -i\hbar \frac{\hat{\bp}_e}{m_e},
\end{eqnarray}
\noindent From this exact relationship, it follows that 
\begin{eqnarray}
    (E_I - E_J) \langle \Phi_I| \hat{\bm r}_e | \Phi_J\rangle= -i\hbar \frac{1}{m_e} \langle  \Phi_I |\hat{\bp}_e |\Phi_J \rangle .\label{eq:rp_eqiv}
\end{eqnarray}

\noindent Then, if we plug Eq. \ref{eq:rp_eqiv} into Eq. \ref{eq:pe_nafie} and use the fact that $\bra{\Phi_{I}} \hat{\bd}^A\ket{\Phi_{I}}= 0$ \cite{Matsika_2007} (so that we can pull out a resolution of the identity), we find:

\begin{eqnarray}
    &&\langle \Psi_I | \hat{\bp}_e | \Psi_I\rangle  \\
    &&=2 m_e  \text{Re}  \sum_{J } \langle \Phi_I |\hat{\br}_e |\Phi_J\rangle\langle \Phi_J |\sum_A \frac{  \bP^A \cdot \hat{\bd}^A}{M_A}  | \Phi_I\rangle  \\
    &&= 2 m_e  \text{Re}   \langle \Phi_I | \hat{\br}_e \sum_A \frac{  \bP^A \cdot \hat{\bd}^A}{M_A}  | \Phi_I \rangle \\
     &&= 2 m_e  \text{Re}   \langle \Phi_I | \hat{\br}_e |\frac{d}{dt}  \Phi_I \rangle \\
      &&= m_e  \frac{d}{dt}  \langle \Phi_I | \hat{\br}_e |\Phi_I \rangle \label{eq:nafie}
\end{eqnarray}

If the math above is tedious, the reader might gain intuition from considering the very simplest case, a hydrogen atom, with Hamiltonian: 

\begin{eqnarray}\label{eq:Hexplicit}
&&\hat{H} = \frac{1}{2M_H}\hat{\bP}^2_H + \frac{1}{2m_e}\hat{\bp}^2_e -  \frac{e^2}{|\bR_H-\br_E|}
\end{eqnarray}

A naive view of BO theory would freeze the nucleus and solve for the electron in that frozen frame. Of course, such a perspective misses the fact that the motion of the nuclei likewise imparts momentum on the electron. One can understand these dynamical correlation effects as capturing the ``electronic inertia''--which are present even for the hydrogen atom.\cite{Kutzelnigg2007} In such a case, the electron inertial effects can be separated simply by going to the center-of-mass frame.

\begin{eqnarray}\label{eq:Hcom}
\hat{H}_{com} =\frac{1}{2M_H + m_e}\hat{\bP}^2_{com} + \frac{1}{2\mu}\hat{\bp}^2_\mu -  \frac{e^2}{|\hat{\br}|}
\end{eqnarray}

The eigenvalues Eq. \ref{eq:Hexplicit} using the BO approximation (assuming frozen nuclei) are of the form $E_n = \frac {P^2} {2M } -\frac{m_e e^4}{8 \epsilon_0^2 h^2} \frac{1}{n^2}$, while the eigenvalues of Eq. \ref{eq:Hcom} are $E_{n}^{\rm com} = \frac {P^2} {2(M + m_{\rm e})}  -\frac{\mu e^4}{8 \epsilon_0^2 h^2} \frac{1}{n^2} $. Thus, in this case, accounting for electronic momentum requires simply replacing the raw electronic mass with the reduced mass $\mu = (M^{-1} +  m_e^{-1})^{-1}$; more interestingly, by isolating the center of mass which is composed of both the nucleus and the electron, $E_{n}^{\rm com}$ includes the overall translation energy of the electron (a result which is lacking in standard BO theory\cite{nafie1983}).

The end conclusion of the above analysis is that, for motion on a single surface, BO erroneously ignores electronic momentum and to recover such momentum, we must go beyond BO theory and work perturbatively with a collection of infinitely many electronic states (that are all weakly coupled to the state of interest). In analogy to electronic structure theory, this is the limit of ``dynamic correlation''. Historically, chemists interested in these dynamics have discussed inertial effects for the hydrogen atom extensively\cite{Kutzelnigg2007}, a concept which has been further explored in the small molecule community.\cite{bunker1977, bubin2013}




\subsection{Conical Intersections And The Limit of Static Correlation}

Now, the argument above was centered around one electronic state being well separated from all other electronic states, and we found that BO fails insofar as $\langle \Phi_I | \hat{\bp}_e| \Phi_I \rangle=0$. That being said, the argument must be adjusted for dynamics near a curve crossing and/or a conical intersection, where it is clear that the total wavefunction will be a complex superposition of wavepackets on different electronic BO states. After all, around a conical intersection (CI), the derivative couplings diverge, driving wavepackets onto multiple surfaces. In such cases, if one propagates a meaningful wavefunction on a collection of BO states, note that the off-diagonal $\langle \Phi_I | \hat{\bp}_e| \Phi_J \rangle \ne 0$ so that one \emph{can} find a non-zero electronic momentum. Indeed, a slew of experimentalists have studied electronic momentum in the vicinity of CI-mediated transitions over the past few decades. \cite{leone2020electronicCI} \\

Nevertheless, the obvious question remains: If BO theory fails to capture the correct electronic momentum on a single surface, can we trust the electronic momentum gathered by dynamics on a few (say, two) strongly coupled electronic states? This limit is known as the ``static limit'' within the electronic structure community. Interestingly, it is known that there is no clear distinction between static and dynamics electron-electron correlation. \cite{Eduard_corr,Marques_corr} This further begs the question of whether of not one can make a clear distinction in the case of nuclear-electron correlation.

In order to address the question of electronic momentum in the static coupling limit, we must first eliminate two sources of confusion.

\subsubsection{Source of Confusion \#1: Berry Phase Effects}

 Ever since the work of Mead and Truhlar \cite{Mead1979} and Berry \cite{berry1984quantal}, it has been well known that a closed path around a CI generates a nonzero phase. Moreover, in the context of a magnetic field to break symmetry, there is a nonzero on diagonal derivative coupling. In recent years, theorists have used the curvature of this phase to calculate a geometric Berry force. If the Berry force is included in the dynamics\cite{Krishna2007,Takatsuka2005}, the nuclear-electronic wave-packet are \emph{total} momentum conserving.\cite{xuezhi:2023:total_ang_bomd,coraline:2024:jcp:ehrenfest_conserve} In other words, it is known that including berry curvature effects is one means to account for electronic momentum.
 Within the physics community, including such a Berry force goes under the title of ``quantum geometry''.\cite{quantum_geometry}\\
 
 Interestingly, however, even though chemists usually consider geometric effects only in the vicinity of a CI\cite{yarkony_diabolical_98} (in the limit of static correlation), the notion of the using a Berry force is valid only valid far away from the CI (in the limit of dynamic correlation). Moreover, we must emphasize that without an external magnetic field (time-reversal symmetry conserving), the on-diagonal Berry force is zero, thus one cannot use the on-diagonal Berry curvature to extract electronic momentum. In short, one certainly cannot invoke on-diagonal Berry curvature as a general means of understanding electronic momentum near a conical intersection.

\subsubsection{Source of Confusion \#2: Electron-Electron Correlation vs. Electron-Nuclear Correlation}

When discussing momentum in the context of static correlation, it is crucial to distinguish between electron-electron correlation and electron-nuclear correlation:

\begin{itemize}
\item As stated above, in electronic structure theory, \emph{static} (or non-dynamic) electron-electron correlation refers to the necessity of mixing two or more Slater determinants whenever they become nearly degenerate in energy. Consider the H\(_2\) molecule. Near equilibrium, a single reference approach can be used, such as Hartree-Fock or Kohn-Sham density functional theory (DFT), to describe the electronic wavefunction. However, as the bond is stretched, the bonding (\(\bigl|\sigma \bar{\sigma}\bigr\rangle\)) and antibonding (\(\bigl|\sigma^{*} \bar{\sigma}^*\bigr\rangle\)) configurations approach degeneracy and the true electronic state takes the form $|\Psi_{\rm el}\rangle \approx \frac{1}{\sqrt{2}}\big(\bigl|\sigma \bar{\sigma} \bigr\rangle + \,\bigl|\sigma^{*} \bar{\sigma}^*\bigr\rangle\big)$. Capturing this static electron–electron correlation requires a multireference treatment such as complete active space self-consistent field (CASSCF) or multireference configuration interaction (MRCI); without a multireference wavefunction, one describes CIs with the incorrect dimensionality of the branching plane.\cite{martinez:2006:ci_topology_wrong,Matsika_2007}

\item For electron–nuclear correlation, dynamics cannot be confined to a single surface, and trajectories must be allowed to either jump between surfaces (e.g. in surface hopping) or move on average surfaces (as in Ehrenfest dynamics).
\end{itemize}


Thus, in order to correctly capture dynamics in the limit of a curve crossing, one must solve two difficult and entangled problems; propagating with good dynamics on a surface generating by single reference electronic structure is not accurate and not sufficient; neither is it sufficient to propagate with BO dynamics on a multireference surface. One cannot extract meaningful electronic momenta without solving two problems at once.
For this reason, in what follows, as we asses electronic momentum in the context of conical intersections, we will necessarily diagonalize an electronic Hamiltonian when possible and avoid using single determinant approximations unless otherwise noted.

\subsection{An Outline of the Paper}


Although the nature of CIs are very well known today\cite{Matsika_2007, yarkony1996, yarkony_diabolical_98}, in this paper, our goal is to reexamine the nature of a CI in a manner whereby we correctly treat the electronic momentum problem. As discussed below, in order to go beyond BO theory, we will invoke a phase space (PS) approach that can (at least partially) resolve the electronic momentum problem. An outline of this paper is as follows. In Sec. \ref{sec:phase_space_background}, we review the theory of phase space electronic Hamiltonians whereby the electronic stationary states are are parameterized by both nuclear position \textbf{R} and momentum \textbf{P}, so that one includes some nonzero electronic momentum (which still conserving the total momentum). 
In Sec. \ref{sec-results}, we examine a model problem with a well-known CI, a beryllium-hydrogen insertion. We examine the nature of the branching plane (which has different dimensionality for BO theory vs PS theory), we study the topology of the PS surface (where we find new stationary points), and we analyze the nature of the resulting eigenstates (where we indeed calculate nonzero electronic momentum and justify the title of this paper). We also propose a novel approach to diabatize PS states near a CI. 
Finally, in Sec. \ref{sec:crhf_instability}, we explore the connection between PS theory and symmetry broken complex restricted Hartree–Fock (CRHF) solutions that naturally emerge at or near conical intersections by driving the system to carry finite electronic momentum. We conclude in Sec. \ref{sec:Conclusions}, where we hypothesize about the dynamical implications of finding stable PS eigenstates with nonzero electronic momentum near curve crossings. 

A word about notation. In this paper, all three (or 3$N_{atom}$) dimensional vectors are written in bold, e.g. $\bR$. Electronic operators are written with a hat, e.g. $\hat{H}$. Electronic states are denoted with subscripts $I,J,K$. Nuclei are indexed by $A,B,C$.

\section{A Brief Review of Phase-Space Electronic Structure Theory}
\label{sec:phase_space_background}

As a brief reminder, one simple practical approach to go beyond BO theory and account for electronic momentum is to use phase space theory. Let us now delineate the relevant theory using the notation which is standard for BO theory.
When describing molecular systems, the standard BO {\em framework} proceeds by initially defining a unitary operator $\hat U$ 
that diagonalizes $\hat H_{\rm el}$ ($\hat U\hat H_{\rm el} \hat U^\dagger = \hat V_{\rm ad} $). One then transforms to the new adiabatic basis,
\begin{eqnarray} \label{eq:Had}
  \hat H_{\rm ad} = \hat U^{\dagger} \hat H \hat U =  \frac {(\hat {\bm P}- i\hbar \hat {\bm d})^2} {2M} + \hat V_{\rm ad}(\hat {\bm R}).
\end{eqnarray} 
where $\bm{d}_{IJ} = \bra{\Phi_I}\frac{\partial}{\partial \bm{R}}\ket{\Phi_J}$ is the derivative coupling vector. The BO {\em approximation} is then that one can neglect the $\hat {\bm P} \cdot \hat{\bm d}$ and $\hat {\bm d} \cdot \hat{\bm d}$ terms in Eq.~\ref{eq:Had}, and diagonalize:
\begin{eqnarray}
\label{eq:HBO:easy}
    \hH_{\rm BO}  = \frac {\hat {\bm P}^2} {2M} + \hat V_{\rm ad}(\hat{\bm R}).
\end{eqnarray}

\noindent where $\hat{\bm{P}}$ acts only on the nuclear wavefunction. As noted above, BO electronic states do not include any nuclear-electronic dynamic correlation with excited states. \\

In order to go beyond BO theory, the phase space approach is to generate electronic states which depend not only on the nuclear coordinates \textbf{R} but also on their momenta \textbf{P}. This flexibility allows the electron to respond as the nuclei move and pick up momentum. Furthermore, if one works with a meaningful phase-space Hamiltonian, one can show that the total momentum—both nuclear and electronic—will be conserved.


Early work by Shenvi\cite{Shenvi2009} proposed an electronic Hamiltonian ($\hat{H}_\mathrm{Shenvi}(\bm{R},\bm{P}) $) that was parametrically dependent on both the nuclear positions $\{\bm{R}_A\}$, as in traditional Born-Oppenheimer theory, and on the nuclear momentum $\{ \bm{P}_A\}$:

\begin{widetext}
\begin{equation}
    \hat{H}_\mathrm{Shenvi}(\bm{R},\bm{P}) = \sum_{A, IJK} \frac{1}{2M_A}(\bP_A\delta_{IJ} - \bd^A_{IJ}) \cdot (\bP_A\delta_{JK} - \bd^A_{JK})\ket{\Phi_I}\bra{\Phi_K} + \sum_I E_I(\{\bm{R}\}) \ket{\Phi_I}\bra{\Phi_I}. \label{eq:Shenvi}
\end{equation}
\end{widetext}

\noindent Here, the eigenvalue $E_I$ satisfies $\hat{H}_{el}\ket{\Phi_I} = E_I\ket{\Phi_I}$ and arises by diagonalizing the standard BO electronic Hamiltonian.\\

Running dynamics along an eigen-surface of $\hat{H}_\mathrm{Shenvi}$ correctly yields a non-zero electronic momentum as well as conserves the total (nuclear+electronic) momentum.
However, these dynamics come with several issues. In the context of the present work, the most egregious of these failures is that the derivative coupling blows up near a curve crossing, as shown by the Hellman-Feynman expression:
\begin{equation}
{\bd}^A_{IJ} = \frac{\langle \Phi_I | \bm\nabla_A \hat{H} | \Phi_
J\rangle}{E_J-E_I}
\end{equation}
Thus, a Shenvi potential energy surface is unstable (and often ill-defined), and certainly not dynamically meaningful near a CI. \\

To avoid the pitfalls of Shenvi's theory, a series of ongoing work has sought to distill the most essential parts of the diagonal derivative coupling into a single electron-operator $\bGamma_A$.\cite{Tao2024_abinitio,Qiu2024_ERF,Tao2024_basisfree} In this case, the total Hamiltonian is written as 
\begin{eqnarray}\label{eq:HPS}
    &&\hat{H}_{\mathrm{PS}}(\bm{R},\bm{P}) = \\\nonumber
    &&\sum_{A}\frac{1}{2M_A}(\bm{P}_A - i\hbar\bGamma_A(\bm{R}))^2 + \hat{H}_{el}(\bm R)
\end{eqnarray}

The form of this $\bGamma = \bGamma^{'} +\bGamma^{"}$ operator can be split into two terms, one that ensures total (nuclear + electronic) linear momentum ($\bGamma^{'})$, and one that ensures total angular momentum($\bGamma^{"}$). One form for $\bGamma^{'}$ and $\bGamma^{''}$ that conserves total momentum is as follows.


\begin{widetext}
\begin{eqnarray}
    \bGamma'_A &=& \frac{-i}{2\hbar}(\theta_A(\hat{\bx})\hat{\bp} +\hat{\bp}\theta_A(\hat{\bx})) \\ 
    \bGamma^{''}_A &=& \sum_B \zeta_{AB}(\bm{R}_A - \bm{R}_B^0)\times (\bK^{-1}_B \hat{\bJ}_B) \\
    \hat{\bJ}_B &=& \frac{-i}{2\hbar}((\hat{\bx}-\bm{R}_B)\times(\theta_B(\hat{\bx})\hat{\bp}) + (\theta_B(\hat{\bx})\hat{\bp})\times(\hat{\bx}-\bm{R}_B) + 2\theta_B(\hat{\bx})\hat{\bm s}) \nonumber\\
    \bm{R}_B^0 &=& \frac{\sum_{A}\zeta_{AB}\bm{R}_B}{\sum_{A}\zeta_{AB}} \nonumber \\
    \bK_B &=& \sum_A\zeta_{AB}(\bm{R}_A\bm{R}_A^{T}-\bm{R}_B^0\bm{R}_B^{0T} - (\bm{R}_A\bm{R}_A^T - \bm{R}_B^{0T}\bm{R}_B^0)\mathcal{I}_3)\nonumber
\end{eqnarray}
\end{widetext}

\noindent For more details, see Refs. \citenum{Tao2024_abinitio,Qiu2024_ERF,Tao2024_basisfree}. In short, $\theta(\hat{\bx})$ defines a position dependent partition of unity among the nuclei, $\zeta_{AB}$ is a nuclear locality term, $\bm{R}_B^0$ defines an atomically partitioned local center of mass, and the tensor $\bm K_B$ can be thought of as an atomically partitioned moment of inertia.\\

Intuitively, $\bGamma_A$ 
``drags" the electronic density in the proximity of atom A towards the direction of $\bP_A$. Note that, unlike the full derivative coupling, $\bGamma_A$ does not involve the full orbital response. Nevertheless, this nuclear-electronic momentum coupling scheme has been shown to capture a non-trivial amount of the total non-adiabatic interaction.\cite{Bian2024_inertial} 
Moreover, while the effect of $\hat{\bm \Gamma}_A$ is smaller than the effect of $\hat{\bm d}_A$, the former never diverges (unlike the latter).
By now, several studies\cite{Duston2024_vcd, Tao2024_abinitio, Tao2024_basisfree} have validated this phase-space approach in the limit of dynamic correlation; below, we will explore the implications of using a phase space theory to study the opposite of limit of static correlation (especially in the vicinity of a CI).

We emphasize that all results below will ignore any and all spin-orbit coupling. While previous results\cite{bradbury2025spin} have shown that nuclear momentum-electronic spin coupling does display strong symmetry breaking for spin multiplet species, our results here focus on whether symmetry breaking occurs without spin couplings.

\section{Results}
\label{sec-results}

\subsection{The BeH$_2$ Model Problem With a Conical Intersection}\label{sec:model_problem}

To better understand the nature of electronic momentum in the context of conical intersection, our problem of choice is the well-known  example of beryllium atom inserted between two hydrogen atoms as studied by Pople\cite{CRHF_stability_pople}. In Fig \ref{fig:gbranch}, we plot
the relevant complete active space configuration interaction(CAS-CI) energies (as well as restricted Hartree Fock (RHF) and complex restricted Hartree Fock (CRHF) energies) along a so-called ``H$_2$ insertion'' 
$\lambda$-coordinate\cite{CRHF_stability_MHG} given as:
\begin{eqnarray}
\text{Be}&:&(0,0,0)\nonumber\\
\text{H}&:&(\lambda,2.540-0.46\lambda,0)\label{eq:x_coord}\\
\text{H}&:&(\lambda,-2.540-0.46\lambda,0)\nonumber
\end{eqnarray}
 At $\lambda$ = 0, we have linear $BeH_2$; as $\lambda$ is increased, the hydrogen atoms are pulled away from the beryllium atom, and closer to one another, reaching $Be + H_2$ at around $\lambda$ = 4.0. For all $\lambda > 0$, the geometries have $C_{2v}$ symmetry. 

\begin{figure}[h]
    \centering
    \includegraphics[width=0.9\linewidth]{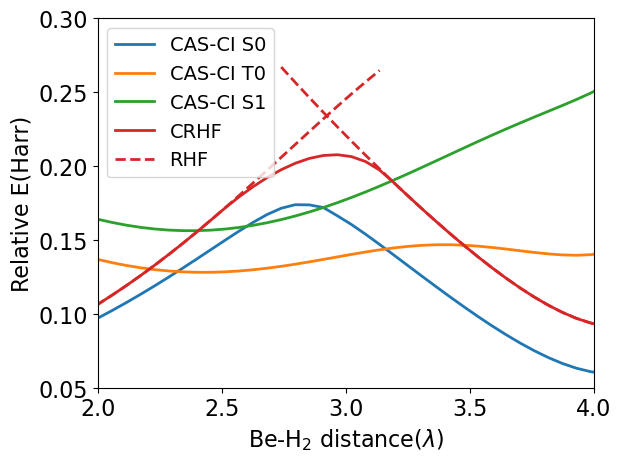}
    \caption{$Be + H_2$ insertion: CRHF and relevant CAS-CI curves; 6-31G basis. State ordering and energies relative to ground state linear H-Be-H. Note that there are two CIs between $S_0$ and $S_1$, at 2.600 and one at 2.902 Bohr respectively. }
    \label{fig:Rscan}
\end{figure}

According to Fig. \ref{fig:Rscan},
there are two singlet-singlet conical intersections along the $\lambda-$direction, one at 2.600 and one at 2.902 Bohr. Our results below focus on the the former CI point, but we believe the results are qualitatively the same for both. There are also singlet-triplet intersections, but here we do not include spin orbit coupling, so that all eigenstates are necessarily spin pure and uncoupled. 
For the data in Fig. \ref{fig:Rscan}, the active space includes all orbitals except the 1s orbital on the beryllium atom (which is frozen).

\subsubsection{The Branching Plane Within BO Theory}\label{sec:plane_condition}

The simplest model for a conical intersection between two singlet states is to assume the existence of a diabatic basis and write: 

\begin{equation}\label{eq:CImat}
\hH = \begin{pmatrix}
H_{11} & H_{12} \\
H_{21} & H_{22} 
\end{pmatrix}
\end{equation}

At a conical intersection seam, the eigenvalues of $\hat{H}$ are degenerate, requiring that $H_{11} - H_{22} = 0$ and $H_{12} = 0$.\cite{Matsika_2007} Let $N_{int}$ be the number of internal degrees of freedom in coordinate space, traditionally 3N-6, where N is the number of atoms and the translations and rotations have been removed. Within these $N_{int}$ degrees of freedom, two conditions must be satisfied to remain on the seam, so that the branching plane has dimension two. In principle, one might require a third condition, since $H_{12}$ may be complex; however, the Hamiltonian for an even number of electrons can always be written in a real-valued fashion as a consequence of time-reversal symmetry.\cite{Mead1979}
Thus, we can write:

\begin{eqnarray}
\hH = E_0(x,y) \mathbb{I} + \begin{pmatrix}
\bm{g} z & \bm{h} x \\
\bm{h}x & -\bm{g} z 
\end{pmatrix}\end{eqnarray}

\begin{eqnarray}
E_{\pm} = E_0(x,y) \pm \sqrt{(\bm{g}x)^2 + (\bm{h}z)^2}
\end{eqnarray}


\noindent where $\bm{g} = \bm\nabla_R (H_{11}-H_{22})$, and $\bm{h} = \text{Re}(\bm\nabla_R H_{12})$. Visual representation of the branching modes for the BO CI for the $BeH_2$ system can be found in Figs. \ref{fig:hbranch}-\ref{fig:fbranch} and have been reported previously.\cite{AlH2yarkony}

\subsubsection{The Branching Plane Within PS Theory}
\label{sec:PS_branch}


Let us now address CIs through the theory of phase space electronic Hamiltonians. As far as the electrons are concerned, the inclusion of the nuclear momentum through the operator $\bP\cdot \bGamma$ breaks the time-reversibility of the Hamiltonian. Therefore, it is no longer possible to write $H_{12} \in \mathbb{R}$, and there exists a third condition $Im(H_{12})=0$. Thus, if we let $N^{int}_{PS} = 6N-6$ be the number of degrees of freedom in phase space, the CI seam must have dimension $N^{int}_{PS} -3$.
Note that $N^{int}_{PS} \ne 6N-12$ because the inclusion of translational and rotation momentum does change the PS energy.


Let the three branching plane directions be defined as: $\{\bm{f},\bm{g},\bm{h}\}$ are $\bm{f} = \text{Im}(\nabla H_{12})$,
$\bm{g} = \nabla (H_{11}-H_{22})$, and
$\bm{h} = \text{Re}(\nabla H_{12})$. \cite{Matsika_2007}
Then, the 2-state Hamiltonian and its eigenvalues are 

\begin{eqnarray}\label{eq:H_expand}\hH = E_0(x,y,z) \mathbb{I} + \begin{pmatrix}
\bm{g} z & \bm{h} x -i\bm{f}y\\
\bm{h}x + i\bm{f}y& -\bm{g} z 
\end{pmatrix}\end{eqnarray}

\begin{eqnarray}
E_{\pm} = E_0(x,y,z) \pm \sqrt{(\bm{g}z)^2 + (\bm{h}x)^2 + (\bm{f}y)^2}
\end{eqnarray}

\begin{figure}[h]
  \centering

  \begin{subfigure}[c]{0.3\textwidth}
    \includegraphics[width=\linewidth]{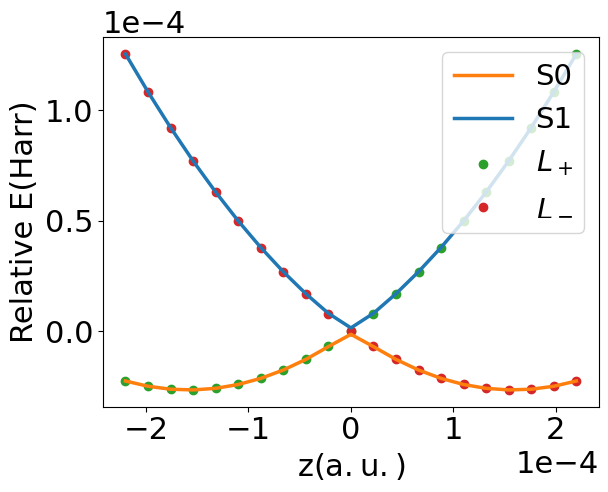}
    \caption{g branch} 
    \label{fig:gbranch}
  \end{subfigure}\hfill
  \begin{subfigure}[c]{0.17\textwidth}
    \includegraphics[width=\linewidth]{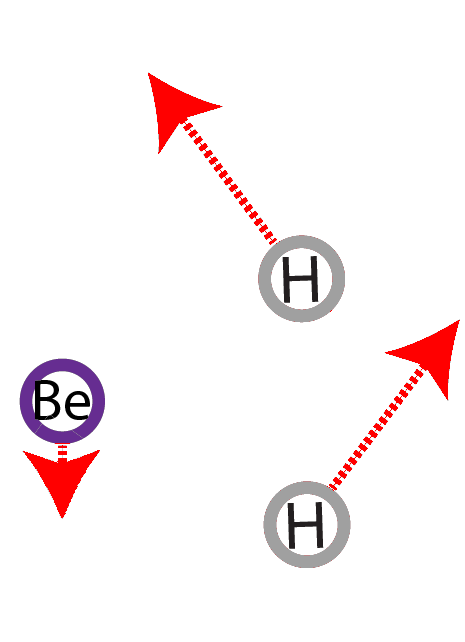}
    \vspace{3em}
  \end{subfigure}
  \begin{subfigure}[c]{0.3\textwidth}
    \includegraphics[width=\linewidth]{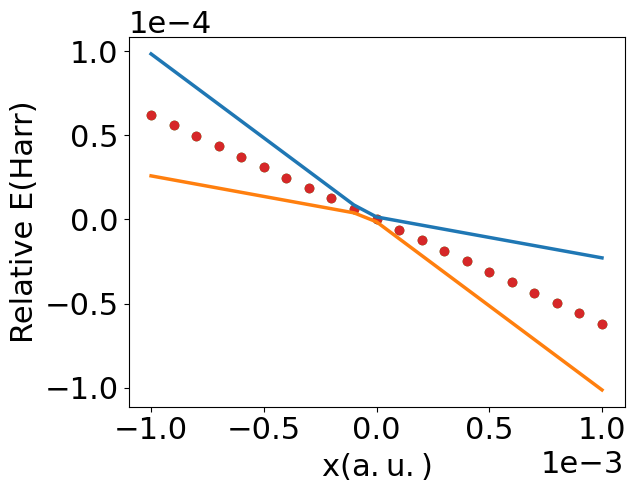}
    \caption{h branch}
    \label{fig:hbranch}
  \end{subfigure}\hfill
  \begin{subfigure}[c]{0.17\textwidth}
    \includegraphics[width=\linewidth]{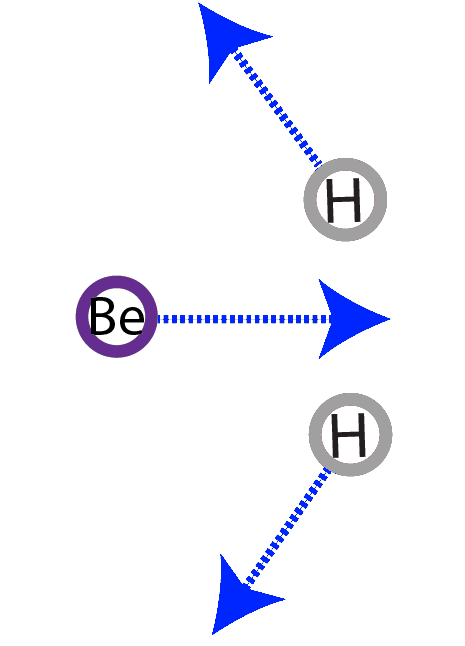}
    \vspace{2em}
  \end{subfigure}


  \begin{subfigure}[c]{0.3\textwidth}
    \includegraphics[width=\linewidth] {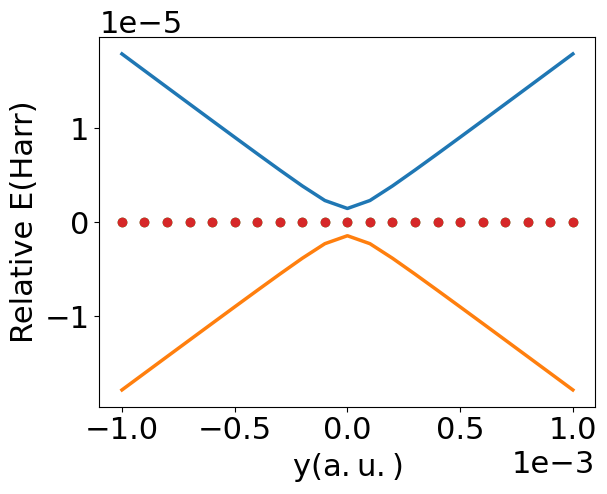}
    \caption{f branch}
    \label{fig:fbranch}
  \end{subfigure}\hfill
  \begin{subfigure}[c]{0.17\textwidth}
    \includegraphics[width=\linewidth]{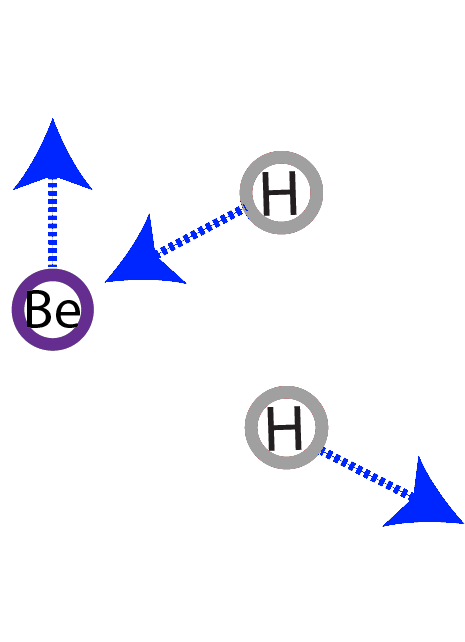}
    \vspace{2em}
  \end{subfigure}


  \caption{Left: CAS-CI and diabatic energies (see Sec. \ref{sec:PS_diabat}) scanned over the branching modes. All energies are relative to the CI point. Note that there are three branching modes. Right: Branching modes for a phase space CI; \textcolor{red}{red} arrows denote component in momentum space, and \textcolor{blue}{blue} arrows denote component in coordinate space.}
  \label{fig:all-modes}
\end{figure}

In Figure \ref{fig:all-modes}, scans of $\{\bm{f},\bm{g},\bm{h}\}$ verify that these modes do indeed break the degeneracy of the adiabatic energies. Note that, because the CI occurs at $\bP = 0$, the branching plane for the PS Hamiltonian exactly incorporates the branching plane for the BO Hamiltonian (at least up to the small $\bGamma^2$ term). 

\subsection{Broken Symmetry Equilibrium States in Momentum around a CI}\label{sec:FCI_topology}

At this point, one important nuance arises. Namely, for a ``conical'' intersection, one requires that the off-diagonal matrix element $H_{12}$ be a linear function of nuclear coordinates (position or momentum). By contrast, if $H_{12}$ were a quadratic function, the intersection would not be conical (it would be a Renner-Teller intersection\cite{yarkony_diabolical_98}). To that end, one interesting question is to scan and explore the PES as a function of nuclear momentum; in fact, the data in Fig. \ref{fig:gbranch} confirms that our intersection is conical. One crucial conclusion from this data is that, when moving along the coordinate direction Fig. \ref{fig:hbranch} and scanning in $\bP$ along the \ref{fig:gbranch} direction, the minima is not always at $\bP = 0$. While such a spin-broken symmetry states has been found previously in the context of spin states, to our knowledge, no such symmetry broken states have yet been found without fine structure.

\begin{figure}[h]
    \centering
    \includegraphics[width=1.0\linewidth]{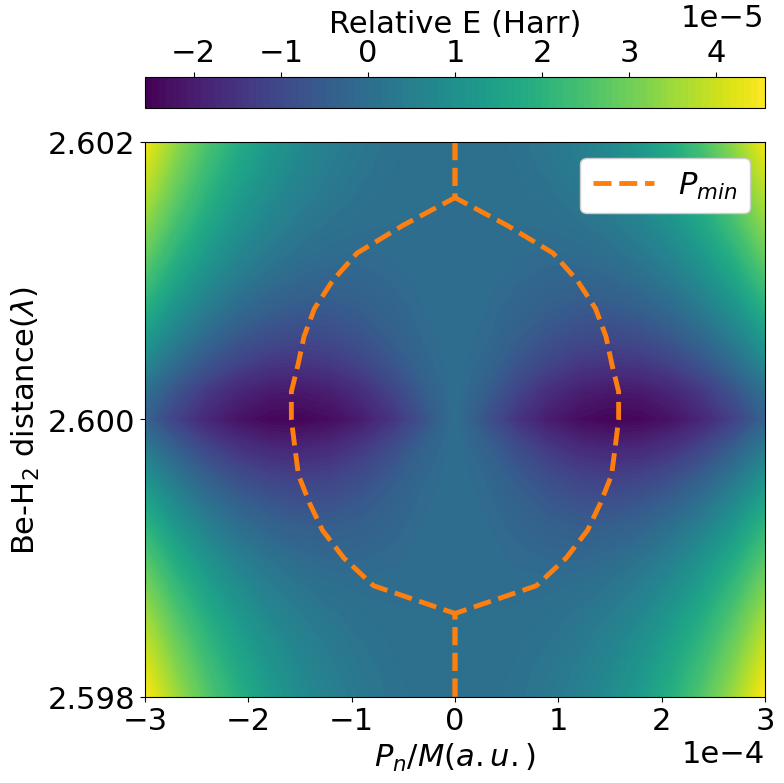}
    \caption{PES scan of the CAS-CI ground state for $BeH_2$. The vertical axis follows the coordinate of Eq. \ref{eq:x_coord}, and the horizontal axis follows the momentum space branching plane depicted in Fig. \ref{fig:gbranch}. For clarity, along every horizontal cut, we subtract the energy at $\bP=0$ (we plot $\tilde{E}(\bR,\bP) = E(\bR,\bP) - E(\bR,0)$). The dashed orange line traces out the location of the minima in $\bP$ as as function of $\bR$. Note that around the CI, there is a double minimum in $\bP$. }
    \label{fig:FCI_adiabat}
\end{figure}

In order to more fully appreciate the topology of the CI surface, in Fig. \ref{fig:FCI_adiabat} we make a two dimensional heat map of the energy as a function of two coordinates: along the $Be+H_2$ insertion coordinate $\lambda$ (Eq. \ref{eq:x_coord}), and along the coordinate of the momentum minima at the CI (depicted in Fig \ref{fig:FCI_adiabat}).
When there are well-separated states (and without spin effects\cite{bradbury2025spin}), there exists only a single minimum in momentum space, located at $\bm{P}=0$. Near the CI, however, we find that the ground and first excited singlet states mix to form a pair of double-well minima.

\subsection{The Electronic Current Density at $P_{min}$}\label{sec:current_density}

\begin{figure}[h]
  \centering
  \begin{subfigure}[c]{0.45\textwidth}
    \includegraphics[width=\linewidth]{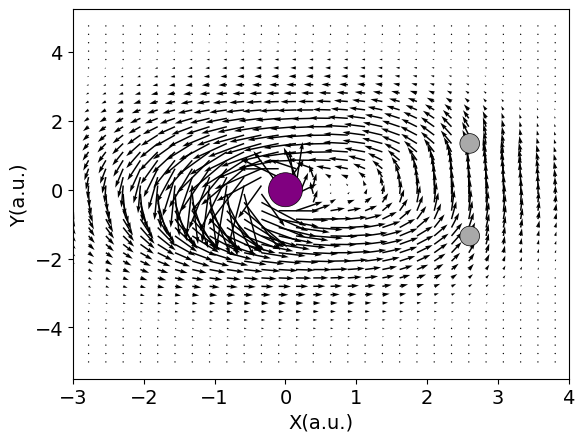}
    \caption{$S_0$ }
    \label{fig:S0_J}
  \end{subfigure}


  \begin{subfigure}[c]{0.45\textwidth}
    \includegraphics[width=\linewidth]{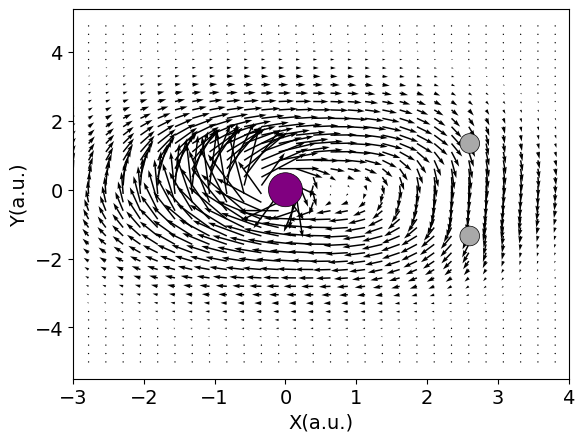}
    \caption{$S_1$ }
    \label{fig:S1_J}
  \end{subfigure}

  \caption{CAS-CI currents at one value of $\bP_{min}$. Including the symmetry breaking operator $\bGamma \cdot \bm P$ leads to  $S_0$ and $S_1$ states  which have currents rotating and counter-rotating with $\bP_{min}$. }

  \label{fig:CASCI_J}

  \end{figure}

To best understand the nature of the stable broken symmetry states at $\bP \ne 0$, let us return to the question of electronic momentum. We have argued that BO theory effectively ignores electronic momentum, while such effects are included in PS theory. One can ask: what is the nature of the electronic motion that is induced at the conical intersection? To answer such a question, in Fig. \ref{fig:CASCI_J}, we plot the current density according to the CAS-CI solutions. Here, the current density at position $\br_i$ is defined as

\begin{equation}\label{eq:Rcurrent}
\langle \hat{j}(\br_i)\rangle = \sum_{pq} \gamma_{pq} \langle \phi_p | \hat{\bp}_e\delta(\br_i) + \delta(\br_i)\hat{\bp}_e | \phi_q \rangle
\end{equation}


\noindent where $\gamma_{pq}$ is the reduced density matrix, and $\{\phi\}$ are a set of molecular orbitals. In second-quantization, for a wavefunction expressed as an expansion of slater determinants ($|\Psi\rangle = \sum_I C_I |\psi_I\rangle$), $\gamma_{pq}$ is given by

\begin{equation}
\gamma_{pq} = \sum_{IJ} C_I^*  C_J \langle \psi_I | \hat{a}_p^\dagger \hat{a}_q | \psi_J \rangle
\end{equation}

In Fig \ref{fig:CASCI_J}, we plot the current elements in the molecular plane after summing over the out-of-plane direction. The emergence of the time-reversal symmetry broken states are rationalized by the following argument: While two states with discordant symmetries do not mix, there is no penalty for such mixing when the states are degenerate. Moreover, if that mixing occurs with a complex phase so as to generate electronic angular momentum, then that electronic angular momentum can couple to nuclear angular momentum through a Coriolis potential, generating lower energy solutions at $\bP \ne 0$.

Future work will necessarily need to address the dynamical implications of these solutions which may have very significant physical consequences. For instance, if a wavepacket passes through a conical intersection but remains on a single adiabatic surface, does the electronic distribution genuinely end up with non-zero electronic momentum? Or more generally, can the presence of CIs “torque” the electronic wavefunction, thus demanding (by momentum conservation) a type of rotational motion in the nuclear or spin wavefunctions? See Sec. \ref{sec:Conclusions} below.


\subsection{Diabatization with $L_z$}
\label{sec:PS_diabat}

Finally, guided by the fact that we find strong, nontrivial electronic momentum for the symmetry-broken states in Figs. \ref{fig:S0_J}-\ref{fig:S1_J}, the last question to ask is: Can we use our new understanding of electronic momentum to further generate the relevant diabatic states? After all, even though exact diabatic states do not exist\cite{Mead1982, yarkony_diabolical_98}, the concept of quasidiabats is powerful and quasidabats should have a reasonably well defined electronic character as a function of configuration space (a fact that allows us to write down the Hamiltonian in Eq. \ref{eq:H_expand} in the first place).

Now, whereas there is a large literature regarding how one may diabatize electronic states using a dipole-based criteria to localize the charge along one part of the molecule, we argue that such an approach may not be optimal for the present situation where the key phenomenon of interest is how electrons acquire angular momentum. To that end, a compelling alternative is to generate a diabatic basis by diagonalizing $\hat{L}_z$, which is motivated by the current plots in Fig. \ref{fig:CASCI_J} demonstrating states with rotating current density. Interestingly, our calculations show that diabatization along $\hat{L}_z$ yields two diabats with effectively the same energies as a function of $\bR$ (though of course the diabatic couplings do depend on $\bR$). 

One final word is appropriate vis a vis this diabatization. Namely, we note that when diagonalizing $\hat{L}_z$, there is an ambiguity as to the center of rotation; the results shown in Fig. \ref{fig:CASCI_J} choose that reference point to be the center of mass, but the results are not very sensitive to the choice of reference.

\section{Discussion: CRHF and The Meaning of Approximate Electronic Structure Symmetry Broken Solutions}\label{sec:crhf_instability}

The results above demonstrate that, for a proper phase space electronic Hamiltonian, the exact eigenstates break symmetry near a conical intersection, leading to two stable solutions at $\bP = \pm \bP_{min}$; the resulting eigenstates carry angular momentum. Interestingly, this finding is consistent with approximate solutions to the BO Schr{\"o}dinger equation at $\bP =0$. In particular,
it has long been recognized that near a conical intersection, a real-valued restricted Hartree–Fock (RHF) solution can become unstable with respect to symmetry-breaking that introduces complex-valued orbitals. Such instabilities, captured within a complex restricted Hartree–Fock (CRHF) theory, arise when an electronic wavefunction is able to lower its energy by allowing previously forbidden orbital mixings and the introduction of electronic states with nonzero electronic momentum. This spontaneous symmetry breaking can occur without external magnetic fields or spin-orbit coupling.\cite{CRHF_stability_MHG,CRHF_stability_pople}

Within the existing literature, the momenta predicted by a CRHF solution is usually interpreted as fictitious; one usually supposes that real/complex symmetry breaking arises only because we seek to minimize the total correlated energy with the unnatural constraint that we are limited to a single slater determinant.\cite{CRHF_stability_MHG, CRHF_stability_pople} Interestingly, one can now ask whether such solutions are in fact physical, when properly compared against a phase space (rather than BO) electronic Hamiltonian that clearly can break symmetry?

Let us now answer this question for the case of BeH$_2$. In Figure \ref{fig:Rscan}, besides the CAS-CI energies, we also plot RHF and CRHF energies. Whereas the RHF energy shows a kink at $\lambda = 2.924$ (near the $S_0$ and $S_1$ CIs) due to the insufficient inclusion of electron-electron correlation, a CRHF solutions does ``smooth'' out the potential energy surface in the vicinity of the RHF kink. Symmetry analysis of the RHF HOMO and LUMO confirm discordant symmetries of $B_2$ and $A_1$. 
Note, however, that such a CRHF solution $\ket{\Psi_{CRHF}}$ is not unique; one can always apply the time-reversal operator to generate $\hat{\mathscr{T}}\ket{ \Psi_{CRHF}}$ which has the same energy but different physical qualities. Moreover, $ \langle \Psi_{CRHF} | \hat{\mathscr{T}}  |\Psi_{CRHF}  \rangle \ne 0$, a fact that might lead one to question the meaning of these states.

\begin{figure}[h]
  \centering
  \begin{subfigure}[c]{0.45\textwidth}
    \includegraphics[width=\linewidth]{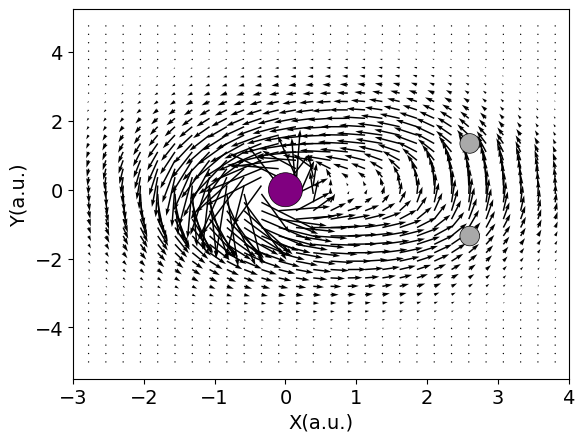}
    \caption{ $\Psi_{CRHF}$ }
    \label{fig:TCRHF_J}
  \end{subfigure}


  \begin{subfigure}[c]{0.45\textwidth}
    \includegraphics[width=\linewidth]{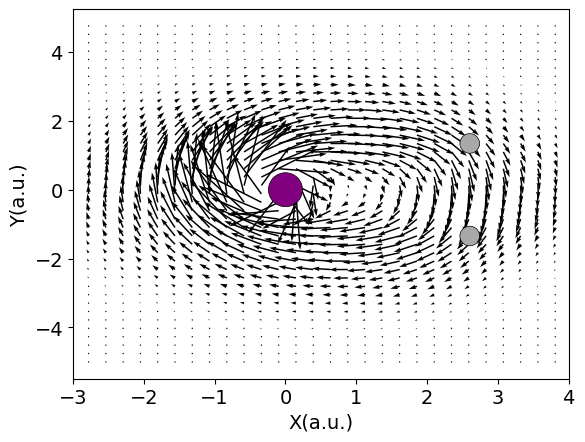}
    \caption{$\mathscr{T} \Psi_{CRHF}$}
    \label{fig:CRHF_J}
  \end{subfigure}

  \caption{ $BeH_2$ CRHF currents at $P_{nuc}=0$. Note the the current closely resemble those of the CAS-CI states (See Fig \ref{fig:CASCI_J})}

  \label{fig:CRHF_J_all}

  \end{figure}

\begin{figure}[H]
    \centering
    \includegraphics[width=0.9\linewidth]{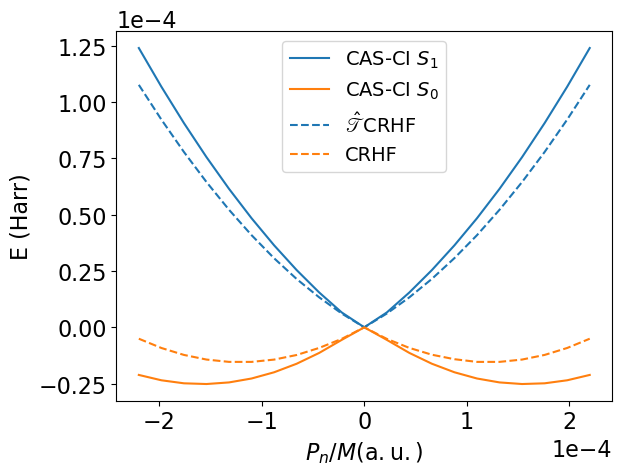}
    \caption{Momentum scan of CRHF, time reversed CRHF, and CAS-CI solutions. Note that the PS CRHF scan  are qualitatively the same as the CAS-CI solutions.}
    \label{fig:CRHF_FCI}
\end{figure}

Interestingly, however, the meaning of these two CRHF solutions (with non-zero electronic momentum) does align with our intuition from phase space theory. Indeed, in Fig. \ref{fig:CRHF_J_all}, we plot the current elements in the molecular plane (inserting the reduced density matrix for the single-reference CRHF determinant into Eq. \ref{eq:Rcurrent}). We find that the BO $(\bP = 0)$ CRHF solution and its time reversed pair correspond to the electronic currents rotating clockwise or counter-clockwise as computed with phase space CAS-CI calculations at $\bm{P} \ne 0$. Evaluating $\langle \hat{L}_z \rangle$, we find that the CRHF solution carries $\approx\pm 0.9\hbar$ quanta of angular momenta. Lastly, in Fig. \ref{fig:CRHF_FCI}, we extend the CRHF solutions from $\bP = 0$ to $\bP \ne 0$ using a phase space electronic Hamiltonian and we show that the CRHF solutions are in fact quite similar the CAS-CI solutions as a function of $\bP$. Thus, within a PS formalism, CRHF solutions do appear to take on some physical meaning.

Of course, the discussion above raises many questions. If we conclude that the CRHF solutions for BeH$_2$ are physically meaningful -- rather than just a compromise that we make to lower energy given the constraints of a limited wavefunction -- one must wonder: does the same conclusion perhaps hold more generally for triplet instabilities\cite{Peach2011} in linear response methods like time dependent (TD)-Hartree Fock or TD-density functional theory (TDDFT) or the random phase approximation (RPA)? More precisely, it is known that imaginary eigenvalues can arise for the linear response ``Casida equations"\cite{casida:1995:book} when the RHF slater determinant of the singlet ground state becomes unstable; such an instability can correspond to a RHF $\to$ UHF instability or a RHF $\to$ CRHF instability (or both).\cite{Peach2011} 
Our data above would suggest that CRHF instabilities will correspond to physically-motivated PS broken symmetry solutions, but future work will necessarily need to investigate the consequences of RHF $\to$ UHF instabilities within phase space electronic structure theory. The last decade has taught us a great deal about the instabilities of response theory\cite{Furche_instability2016, ou:2015:dc_response_theory, herbert:2014:jcp_dercouple, liu:2014:dercouple_div} within the context of a BO Hamiltonian, and much of that knowledge should be transferrable to PS theory.

\section{Conclusions}\label{sec:Conclusions}
In this work we have revisited conical intersections from the perspective of phase-space based electronic structure theory, in which the nuclear configuration space is extended from position ($\bR$), to the full phase space 
($\bR,\bP$). While a conical intersection in the conventional Born–Oppenheimer picture is defined by a two-dimensional branching plane in coordinate space, incorporating nuclear momenta expands the branching sub-space to three dimensions. This extra dimension has a profound consequence in that one can find stationary electronic states at $\bP \ne 0$, so that the adiabatic electronic states take on finite electronic momentum (in this case, electronic orbital angular momentum) when the system approaches the intersection seam. Interestingly, this result can be found not only with CAS-CI calculations but also with much less expensive complex-restricted Hartree–Fock calculations. Hence practical questions about electronic angular momentum, its coupling to nuclear motion, and its role in non-adiabatic chemistry can be probed without invoking the most demanding many-body methods.

Looking ahead, several intertwined problems emerge. First, we have not addressed in any meaningful sense the Berry curvature that arises for a phase space surface.  In particular, note that within BO theory, for a system with an even number of electrons, the electronic wavefunction is always real and the abelian Berry curvature $\Omega_{\bR \bR}$ is always zero -- except at a CI where it is undefined.  By contrast, for a phase space potential energy surface at $\bP \ne 0$, one can imagine constructing the analogous Berry curvature matrix $[\Omega_{\bR \bR} \;   \Omega_{\bR \bP} ;   \Omega_{\bP \bR}  \; \Omega_{\bP \bP}]$ which should almost never vanish.  If $\Omega_{\bR \bR}$ is an effective magnetic field for classical nuclear motion \cite{berry:1993:royal:half_classical}, one must wonder: what is the physical meaning of the phase space analogue? 


Second, although we have not addressed spin effects in this paper, it is very possible that the results above will have bearing on the chiral‐induced spin selectivity (CISS) effect. After all, for systems with spin degrees of freedom, it is the total molecular angular momentum
\[
\mathbf{J}
= \mathbf{L}_{\text{el}}
+ \mathbf{L}_{\text{nuc}}
+ \mathbf{S}
\]
that is conserved. Therefore, 
if motion along one phase space potential energy surface forces the electronic wavefunction to acquire or lose \emph{orbital}
angular momentum,
\(\mathbf{L}_{\text{el}}\), then 
any change
\(\Delta\mathbf{L}_{\text{el}}\neq 0\) must be compensated by either by the \emph{nuclear} component,
\(\Delta\mathbf{L}_{\text{nuc}}\),
or the \emph{spin} angular momentum,
\(\Delta\mathbf{S}\).
In practice, part of the induced torque is absorbed by nuclear torsion, while spin–orbit interactions (as well as the spin Coriolis interaction \cite{bradbury2025spin}) channel the
remaining orbital flow into spin angular momentum. A conical intersection could therefore acts as a conduit that redistributes not just energy but also angular momentum among
electrons, nuclei, and spins--pumping net spin polarization
into the system.\cite{yanze_spin_ci} One could check such a prediction either by running semiclassical phase space surface-hopping simulations (as opposed to standard surface hopping with BO surfaces) that conserve angular momentum\cite{Bian2024} or propagating wavepacket dynamics on properly properly Weyl-transformed surfaces. \cite{xinchun2025_complete} 

Third and finally, one must wonder how the CIs described above will be modified in the presence of external magnetic fields and can they be used to generate magnetic field effects?
\cite{bhati_magnetic_part1, bhati_magnetic_part2, steiner:1989:magnetic_review} Indeed, very interesting questions await us.



\bibliography{Manuscript/reference}

\end{document}